\documentclass[twocolumn,conference]{IEEEtran}
\usepackage{amsmath,amssymb,amsfonts}
\usepackage{algorithmic}
\usepackage{graphicx}
\usepackage{textcomp}
\usepackage{xcolor}
\usepackage{comment}
\usepackage{mathtools}
\usepackage{support-caption}
\usepackage{subfig}
\usepackage{caption}
\usepackage{dblfloatfix}
\usepackage{siunitx}
\usepackage{multirow}

\newcommand{\ddx}[2]{\frac{\partial #1}{\partial #2}}

\newcommand{\mR}{\mathbb{R}}
\newcommand{\mbu}{\mathbf{u}}
\newcommand{\mbuh}{\mathbf{u}^h}

\newcommand{\mbvh}{\mathbf{v}^h}
\newcommand{\mbn}{\mathbf{\hat{n}}}

\newcommand{\p}{p}
\newcommand{\ph}{p^h}

\newcommand{\qh}{q^h}
\newcommand{\mbg}{\mathbf{g}}
\newcommand{\mbnull}{\mathbf{0}}

\newcommand{\romdim}{N_r}
\newcommand{\mbur}{\hat{\mathbf{u}}}
\newcommand{\mbV}{\mathbf{V}}
\newcommand{\deimdim}{N_m}

\newfont{\amsbold}{msbm10}
\newfont{\logobold}{logobf10 scaled\magstep2}

\def\*#1{\mathbf{#1}}

\newcommand{\Reynolds}{\mathit{Re}}

\newcommand{\ba}{\mathbf{a}}

\newcommand{\bu}{\mathbf{u}}

\newcommand{\bx}{\boldsymbol{x}}

\newcommand{\testFunc}{w^h}
\newcommand{\up}{c^{n+1}}
\newcommand{\un}{c^n}
\newcommand{\dt}{\Delta t}

\makeatletter 
\newcommand{\linebreakand}{%
  \end{@IEEEauthorhalign}
  \hfill\mbox{}\par
  \mbox{}\hfill\begin{@IEEEauthorhalign}
}
\makeatother 

\usepackage[unicode=true,
bookmarks=true,bookmarksnumbered=true,bookmarksopen=true,bookmarksopenlevel=1,
breaklinks=false,pdfborder={0 0 0},pdfborderstyle={},backref=false,colorlinks=false]
{hyperref}
\hypersetup{pdftitle={Your Title},
 pdfauthor={Your Name},
 pdfpagelayout=OneColumn, pdfnewwindow=true, pdfstartview=XYZ, plainpages=false}

\makeatletter

\graphicspath{{figures}}

\begin{document}

\title{Towards Real-Time Urban Physics Simulations\\ with Digital Twins

\thanks{Identify applicable funding agency here. If none, delete this.}
}

\author{\IEEEauthorblockN{
Jacopo Bonari,\IEEEauthorrefmark{1}
Lisa K\"{u}hn,\IEEEauthorrefmark{1}
Max von Danwitz,\IEEEauthorrefmark{1}
Alexander Popp,\IEEEauthorrefmark{1}\IEEEauthorrefmark{2}}\\
\IEEEauthorblockA{\IEEEauthorrefmark{1}German Aerospace Center, 
Institute for the Protection of Terrestrial Infrastructures \\
Rathausallee 12, 53757 Sankt Augustin, Germany}
\IEEEauthorblockA{\IEEEauthorrefmark{2}University of the Bundeswehr Munich,
Institute for Mathematics and Computer-Based Simulation \\
Werner-Heisenberg-Weg 39, 85577 Neubiberg, Germany}}

\maketitle

\begin{abstract}
    Urban populations continue to grow, highlighting the critical need
    to safeguard civilians against potential disruptions, such as
    dangerous gas contaminant dispersion. The digital twin (DT)
    framework offers promise in analyzing and predicting such events.
    This study presents a computational framework for modelling airborne
    contaminant dispersion in built environments. Leveraging automatic
    generation of computational domains and solution processes, the
    proposed framework solves the underlying physical model equations
    with the finite element method (FEM) for numerical solutions. Model
    order reduction  (MOR) methods are investigated to enhance
    computational efficiency without compromising accuracy. The study
    outlines the automatic model generation process, the details of the
    employed model, and the future perspectives for the realization of a
    DT. Throughout this research, the aim is to develop a reliable
    predictive model combining physics and data in a hybrid DT to
    provide informed real-time support within evacuation scenarios.
\end{abstract}


\section{Introduction}
Critical infrastructures constitute the backbone of modern societies
and, as cities expand and the number of urban inhabitants grows,
protecting civil population against disruptions of potentially dangerous
systems is a matter of primary importance. Among several different
threats posed to the community by possible incidents, the protection
against dangerous gas contaminant dispersion emerges as a focal point,
given its potential for catastrophic consequences on public health,
environment, and economic activity. At the same time, the conceptual and
practical development of the digital twin (DT) framework has allowed for
it to emerge as a prominent tool of analysis and prediction. Even though
no unique definition of a DT is acknowledged by researchers, the
thorough review by Boyes and Watson~\cite{boyes_digital_2022} finds
in~\cite{eyre_untangling_2020} a comprehensive definition as "\emph{A
live digital coupling of the state of a physical asset or process to a
virtual representation with a functional output}".

In~\cite{brucherseifer_digital_2021}, the DT concept is customized for
the infrastructure domain. Here, a requirements analysis is performed
with the target of improving possible crisis management and defining
measures to increase the resilience of infrastructure components. The
authors conclude that the DT paradigm is appropriate for the scope.
Concurrently, many efforts have been undertaken by researchers to apply
this concept to the analysis of dangerous gas or generic contaminant
dispersion, with particular focus on risks directly connected to oil and
gas industry~\cite{abdrakhmanova_review_2020} and contaminant spread in
small scale closed environments~\cite{nagarajan_predicting_2023}. On the
other hand, to the best of the authors' knowledge, less attention has
been devoted to the analysis of contaminant dispersion in urban or, in
general, built environments. This is in spite of the existence of a
well-established methodology in both computational fluid dynamics (CFD)
applied to urban physics~\cite{blocken_computational_2015} and research
on DT for cities, both on the concept level and applied to specific case
studies~\cite{shahat_city_2021,caprari_digital_2022}.

In this study, we propose a computational framework for the analysis of
airborne contaminant dispersion in a built environment to enhance
situational awareness in a specific risk scenario. The framework
strongly relies on an automatic generation of the computational domain
and solution process. Information about the position and the geometry of
buildings influencing the gas diffusion process can be either collected
automatically according to location based queries, or prompted by a user
in a structured file, while simulation parameters are prompted once at
the beginning of the workflow. The objective is to develop a reliable
predictive model for a DT, aiming to offer informed real-time support
within, for example, evacuation scenarios.

The core part of the workflow foresees the solution of two sets of
partial differential equations (PDEs). The first simulation step deals
with the evaluation of a wind vector field, coincident with the solution
of the incompressible Navier-Stokes (INS) equations; the second step
deals with the atmospheric diffusion of the contaminant agent based on
an advection-diffusion (AD) process, governed by the previously
evaluated wind field. 

The finite element method (FEM) is employed for the numerical solution
of the PDEs. Thanks to its versatility and to the increase of available
computational resources, numerical methods have proven themselves as
reliable instruments in the context under examination. 

During the last three decades, simulation of wind distribution in urban
environments has strongly relied on computational fluid dynamics (CFD).
From the first seminal analyses of wind flows over simple
buildings~\cite{murakami_three-dimensional_1989}, the attention evolved
towards specific studies of pedestrian comfort in windy
environments~\cite{tominaga_aij_2008} and to cases tailored for the
analysis of wind interaction with real
structures~\cite{van_hooff_coupled_2010}.

Since a DT model requires by definition a continuous and bidirectional
stream of data and information, and since it strives to virtually
reproduce a real phenomenon, the capability of delivering real-time
results is of paramount importance. To obtain this goal, model order
reduction (MOR) methods are investigated in the current study to deliver
fast yet accurate results in terms of wind field evaluation, which is
the most computationally intense part of the workflow. To achieve this,
we apply proper orthogonal decomposition (POD) in combination with an
approximation of the nonlinearities of the INS equations following the
works in~\cite{Ballarin.2015,Cicci.2022}. For a general overview of MOR
for CFD in general we refer to \cite{Rozza.2023}. 

The reminder of the article is structured as follows: in
Sec.~\ref{sec:method} a brief overview of the complete workflow is
given, while Sec.~\ref{sec:data_collection},
Sec.~\ref{sec:virtual_replica}, and Sec.~\ref{sec:simrun} present the
methodological details of data collection, preparing a virtual replica
and performing urban physics simulations, respectively.
Section~\ref{sec:application} showcases exemplary solutions for two
different test cases characterized by increasing levels of complexity.
Finally, Sec.~\ref{sec:conclusion} presents future perspectives and
possible model extensions, together with the steps required for a future
migration from a numerical model to a fully bidirectional DT.

\section{Workflow Overview}\label{sec:method}

\begin{figure}
\centering
\includegraphics[width=\columnwidth]{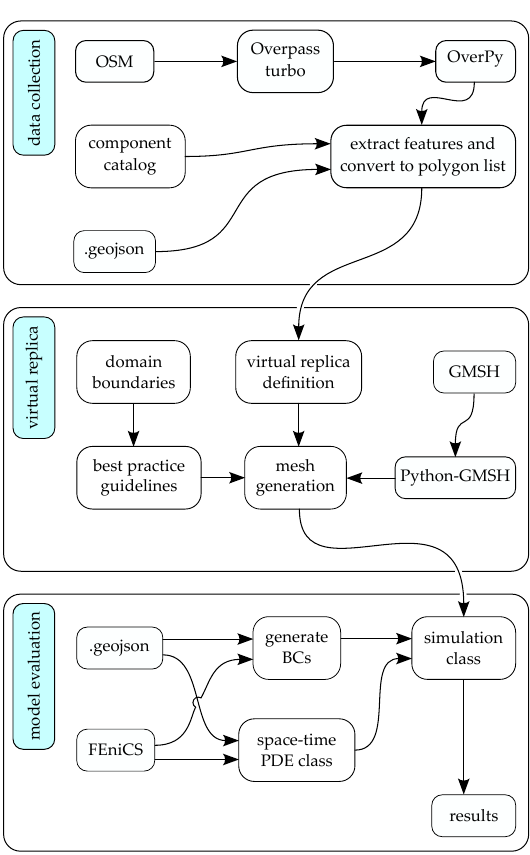}
\caption{Urban physics simulation workflow. Starting from geo-referenced
building data in a database, here OpenStreetMap (OSM), the highly
automatized workflow generates a FEM-mesh using
GMSH~\cite{remacle_gmsh_2009} and performs  simulations based on the PDE
solver FEniCS~\cite{AlnaesEtal2015}.}
\label{fig:workflowMax}
\end{figure}

To enable urban physics simulations for general built environments, we
develop a highly automatized workflow, see Fig.~\ref{fig:workflowMax}.
The complete process is governed by a Python class. The process starts
from geo-referenced building data in a database and delivers at the end
of each simulation run results in a format compliant with standard
geographic information systems (GIS). Details of core ingredients are
outlined in the following sections.  

\section{Domain definition and data
Collection}\label{sec:data_collection} To facilitate a standardized data
collection procedure, we start with a precise definition of the
computational domain that is considered in the urban physics
simulations.

\subsection{Description of the simulation domain}
\begin{figure}
\centering
\includegraphics[width=\columnwidth]{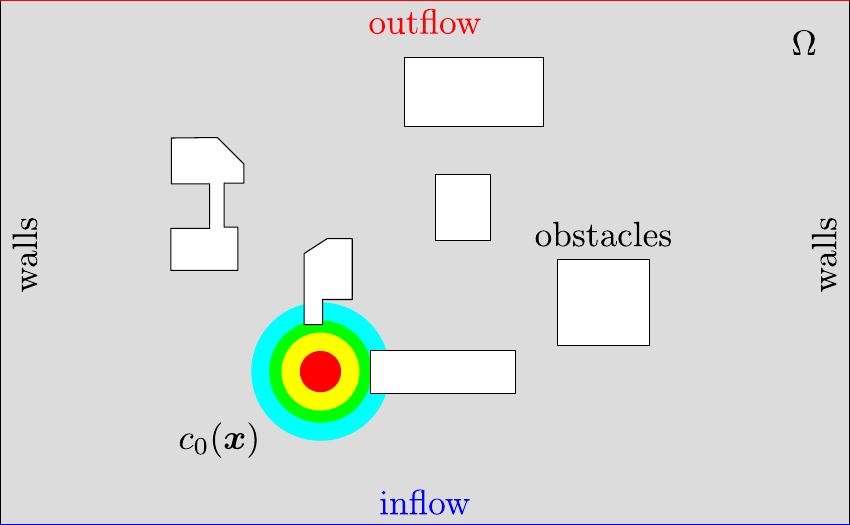}
\caption{Computational domain with highlighted inflow and outflow boundaries and location of initial gas source.}
\label{fig:domain}
\end{figure}

A computational domain $Q_T = (0,t_f)\times\Omega$ is defined with
$(0,t_f)$ an analysis time window and $\Omega$ a rectangular,  connected
subset of $\mathbb{R}^2$. Since the focus of the study hinges on the
distribution of the contaminant in the open air, buildings are
characterized by their outer perimeter only, which represents an
obstacle for free air circulation. Wind is supposed to enter the domain
at constant speed and perpendicular to the bottom side of the model,
while wind velocity is assumed to be null in correspondence of the
lateral sides of the domain and by the buildings sides, the latter being
the only physical boundaries of the domain. Finally, the top is supposed
to be the outflow side, for which free-flow conditions are foreseen. In
symbols: $\partial\Omega = \Gamma_\mathrm{i}\cap
\Gamma_\mathrm{o}\cap\Gamma_\mathrm{n}$, where $i$, $o$, and $n$ are the
subscripts for the \emph{inflow} side, the \emph{outflow} side, and the
\emph{no slip} sides. As a last step, a source for the contaminant leak
is supposed to be located in the surrounding of a specific point
$\bx_\mathrm{c} \in \Omega$ in the form of a scalar field $c_0(\bx)$. A
sketch of the aforementioned environment is depicted in
Fig.~\ref{fig:domain}.

\subsection{Geo-referenced data collection}
An automatic domain construction is then performed exploiting
geo-referenced data according to three different procedures.
\begin{itemize}
    \item Information on building geometries can be recollected using
    OpenStreetMap (OSM)~\cite{OpenStreetMap}, which constitutes a free
    and openly accessible geographic repository, continuously updated
    and managed by volunteers through open cooperation. Within OSM, data
    is collected through public surveys, delineated from aerial visuals,
    or integrated from other freely licensed geographic data outlets.
    Web-based data mining tools for OSM, e.g. Overpass
    turbo~\cite{Raifer2012}, can be used to extract information in a
    format suitable for a seamless conversion to CAD software kernels
    or, in general, geometry libraries. Furthermore, the Python wrapper
    OverPy is employed to access Overpass turbo, making the whole
    geometry acquisition step self-contained within a Python environment
    workflow.
    \item Building locations can also be accessed through a
    \emph{component catalog}~\cite{Stuermer2023}, consisting of a
    database containing datasheets, components, symbols, and code in a
    graph-shaped structure that facilitates the integration of
    components into the proposed simulation environment.
    \item Finally, data can also be prompted providing a custom
    \texttt{.geojson} file, in which every feature describes a building
    in terms of type and geometry, the latter being limited for now to
    the list of its corners according to an established geodesic
    reference system.
\end{itemize}

\section{Virtual replica}\label{sec:virtual_replica} With the polygons
extracted, an overall rectangular domain border can be defined, where
the clearance of each side from the buildings cluster can be defined
according to different strategies. These borders do not represent a
physical boundary, and, to avoid the introduction of spurious effects
during the simulation, they must be located far enough from the
represented physical objects. Here, the concept of \emph{blockage ratio}
($BR$)~\cite{blocken_computational_2015} is applied, which states that
artificial accelerations are avoided when the ratio between the
projected length of the buildings' sides on the domain border
perpendicular to wind direction, and the border itself, is less than
$17\%$.

\subsection{Mesh generation}
\begin{figure}
\centering
\includegraphics[width=\columnwidth]{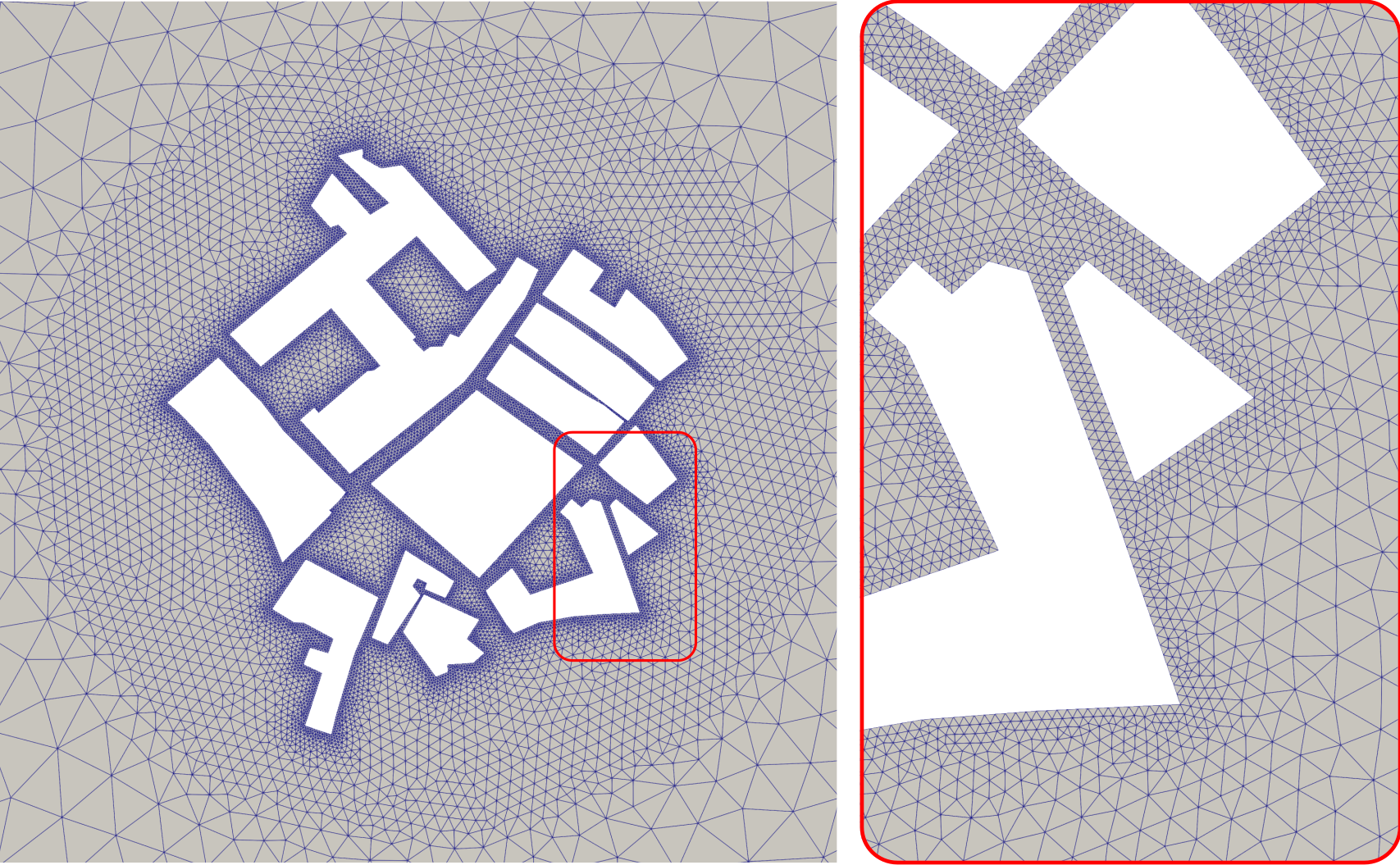}
\caption{Analysis-ready FEM mesh with automatic local refinement in the narrow crevices between the buildings.}
\label{fig:mesh}
\end{figure}

After the two-dimensional domain is defined as a surface in terms of
bounding polygons, the geometry is discretized into a FEM mesh by the
GMSH~\cite{remacle_gmsh_2009} Python API Python-GMSH, with proper
physical tags assigned to discriminate among portions of the boundary
representing inflow and outflow portions, and the no-slip sides, where
the wind velocity is supposed to be zero. The whole mesh generation
procedure is carried out automatically provided that characteristic mesh
sizes $l_{\mathrm{c}}$ are defined during the initialization step. This
is necessary in order to satisfy the convergence criteria of the
numerical solution procedure of the PDEs describing the problem. In the
specific example, three different values of $l_{\mathrm{c}}$ have been
employed, each resulting in a increasing level of refinement,
Fig.~\ref{fig:mesh}. A very fine mesh is defined on the sides of the
buildings, a requirement demanded by an accurate wind field evaluation;
a mesh with slightly larger elements is then defined between the
buildings, small enough to guarantee the stability of the numerical
evaluation of the contaminant dispersion; finally, a coarse mesh is
constructed in the area far from the buildings, since this area only
represents a buffer zone necessary to respect the requirement on the
$\mathit{BR}$ parameter. The triangular mesh is generated with a frontal
Delaunay algorithm allowing for automatic local mesh
refinement~\cite{Rebay.1993}.

\subsection{Boundary and initial conditions}
The last input requirement consists in the definition of boundary and
initial conditions required by the two sets of PDEs. Concerning the INS
problem, the required input data are \emph{wind intensity} and
\emph{wind direction}; regarding the AD problem they consist in an
initial distribution of the contaminant under examination. Since the
final goal of our research foresees the integration of the workflow into
a DT framework, this data is supposed to be continuously collected
through sensors, e.g. metereological stations and detection control
units. At the current preliminary stage, a structured \texttt{.geojson}
file is hard-coded at the beginning of the process and prompts to the
required pieces of information.

\section{Physics-based Simulation}\label{sec:simrun} The core section of
the simulation run consists in the subsequent numerical solution of the
INS and AD systems of equations, which are both performed using
FEniCS~\cite{AlnaesEtal2015}, an open-source computing platform for the
solution of PDEs. The two sets of equations are coupled through the wind
field, which is first obtained as solution of the INS problem, and then
considered as a background field during the AD analysis. Given the
highly nonlinear nature of the INS equations, this is the most
computationally intensive step. Since it can be reasonably assumed that
the variation of the underlying meteorological conditions happens on a
time scale higher than the one related to the diffusion process, the
wind field is considered to be steady and therefore evaluated only once
at the beginning of the procedure. Furthermore, MOR techniques are applied to further speed up the wind field
evaluation, striving to achieve real-time simulation runs in the medium
and long term. On the other hand, the AD process can be described by a
linear system of equations, thus making its solution much less
computationally expensive.

\subsection{Simulation of Wind Flow Field}\label{FOM}

\subsubsection{Full-Order Model}
We consider the steady-state incompressible Navier-Stokes equation
system~\cite{Elman.2014}. Written in terms of the kinematic viscosity
$\nu$, it reads:
\begin{equation}\label{eq:INS_strong}
    \begin{aligned}
        -\nu \nabla^2\mbu + \mbu\cdot\nabla\mbu + \nabla \p &= \mbnull \quad &\text{in} \quad &\Omega, \\
        \nabla \cdot \mbu &= 0 \quad &\text{in} \quad &\Omega, \\
        \mbu &= \mbg(\mu) \quad &\text{on} \quad &\Gamma_D, \\
        \nu \frac{\partial \mbu}{\partial \mbn} - \p \mbn &= \mbnull \quad &\text{on} \quad &\Gamma_N,
        \end{aligned}
\end{equation}
where $\mbu$ denotes the wind velocity and $\p$ the pressure. The
boundary conditions are moreover defined on the Dirichlet boundary
$\Gamma_D$ and the Neumann boundary $\Gamma_N$. To account for different
inflow velocities, a multiplicative factor $\mu \in \mR$ is considered
in the inhomogeneous Dirichlet boundary condition, so that
$\left(\mbu,\p\right) =
\bigl(\mbu(\mu),\p(\mu)\bigl)$~\cite{Ballarin.2015}. Moreover, we refer
to the maximum Reynolds number as $\Reynolds =
\|\mbu_{\mathrm{max}}\|l/\nu$, where $\|\mbu_{\mathrm{max}}\|$ is the
maximum value of the velocity magnitude and $l$ is the characteristic
length of the domain. 

To facilitate a numerical solution of the wind field, the strong form of
the Navier-Stokes equation system described by Eq.~\eqref{eq:INS_strong}
is transferred into a weak form and discretized using Taylor-Hood finite
elements~\cite{Elman.2014}. The parametric, inhomogeneous Dirichlet
boundary condition accounting for varying wind speeds $\mbg(\mu)$ is
treated with a so-called lifting function~\cite{Key.2023}. For the sake
of brevity, we only state a short version of the resulting discretized
weak form and refer to the literature for a detailed derivation and
reference implementation~\cite{Ballarin.2015,RBniCS_website}:
\begin{align}\label{eq:ins_fom}
        a(\mbuh,\mbvh;\mu) + b(\mbvh,\ph;\mu) 
        + c(\mbuh,\mbuh,\mbvh;\mu) &= 0\\ \nonumber
    b(\mbuh,\qh;\mu) &= 0.
\end{align}

The linearized form of the problem is finally solved with the Newton
algorithm implemented within the FEniCS framework~\cite{FEniCS_book}.

\subsection{Simulation of Contaminant Transport}
The airborne contaminant transport is modeled with the following
time-dependent advection-diffusion equation:
\begin{equation}
\mathcal{R}(c) \coloneqq  \ddx{c}{t} + \bu \cdot \nabla c - k \, \Delta c = 0. \label{eq:ADstrong}
\end{equation}
Therein, the scalar unknown $c(\bx, t)$ represents the contaminant
concentration as a function of the spatial coordinates $\bx$ and time
$t$. The advection velocity is a given vector field $\bu(\bx)$, obtained
as solution of Eq.~\eqref{eq:INS_strong} and in practice approximated
with $\mbuh(\mu) \approx \mbV\mbur(\mu)$. Furthermore, the constant
diffusion coefficient is denoted by $k$ and the Laplacian operator can
be expressed as $\Delta(\cdot) = \nabla \cdot \nabla (\cdot)$, based on
the spatial gradient operator $\nabla (\cdot)$.

The resulting initial boundary value problem (IBVP) is defined on the
already familiar domain $\Omega$ shown in Fig.~\ref{fig:domain}.
Moreover, we denote the derivative of $c$ in the direction of the
outward-facing boundary normal as $\partial c/\partial
\mathbf{\hat{n}}$. The IBVP states that we require
Eq.~\eqref{eq:ADstrong} to hold on $\Omega$, along with a known initial
concentration distribution $c_0(\bx)$ and given Dirichlet boundary
condition $g$. The complete IBVP reads:
\begin{align}
\label{eq:IBVP}
\text{IBVP} \quad
    \begin{cases}
    	\mathcal{R}\bigl(c(\bx, t)\bigl) = 0, &\;  \text{on} \; \Omega,  \\
    	c(\bx, t) = c_0(\bx),& \;  \text{at} \; t=0,   \\
    	c(\bx, t) = g(\bx, t), &\;  \text{on} \; \Gamma^\mathrm{D}, \\
    	\partial c/\partial \mathbf{\hat{n}}= 0, &\;  \text{on} \; \Gamma^\mathrm{N}. \\ 
    \end{cases}
\end{align}

With suitable finite element function spaces, the well established
SUPG-stabilized weak form of the boundary value problem can be derived
following~\cite{Brooks.1982, Behr.2008, KKey.2023}. The transient nature
of the problem is treated with an implicit Euler time stepping
algorithm~\cite[Equation (10.25)]{Elman.2014}, which results in the
following discretized weak form
\begin{alignat}{2}
	\label{eq:weakAD}
	& & & \int \testFunc \cdot \up \, d\Omega \\ \nonumber &+& & \dt
	\int \testFunc \cdot \left( \bu \cdot \nabla \up \right) \, d\Omega
	\\ \nonumber &+& & \dt \int \nabla \testFunc \cdot \left( k \nabla
	\up \right) \, d\Omega \\ \nonumber &+& & \int \ba \cdot \nabla
	\testFunc \cdot \tau \cdot \up \, d\Omega \\ \nonumber &+& & \dt
	\int \bu \cdot \nabla \testFunc \cdot \tau \cdot \left( \bu \cdot
	\nabla \up - \nabla \cdot k \nabla \up  \right) \, d\Omega \\
	\nonumber &=& & \int \testFunc \cdot \un \, d\Omega + \int \bu \cdot
	\nabla \testFunc \cdot \tau \cdot \un \, d\Omega.
\end{alignat}

After an initial assembly of the linear system of equation,
time-stepping only requires an update of the right hand side and the
solution of the resulting system.

\section{Application Examples: Munich and D\"usseldorf}\label{sec:application}

\begin{figure*}
	\centering
    \subfloat[][Exemplary wind field of Geometry 1 for $\Reynolds\approx10$.\label{subfig:wind_field_campus}]
	{\includegraphics[width=\columnwidth]{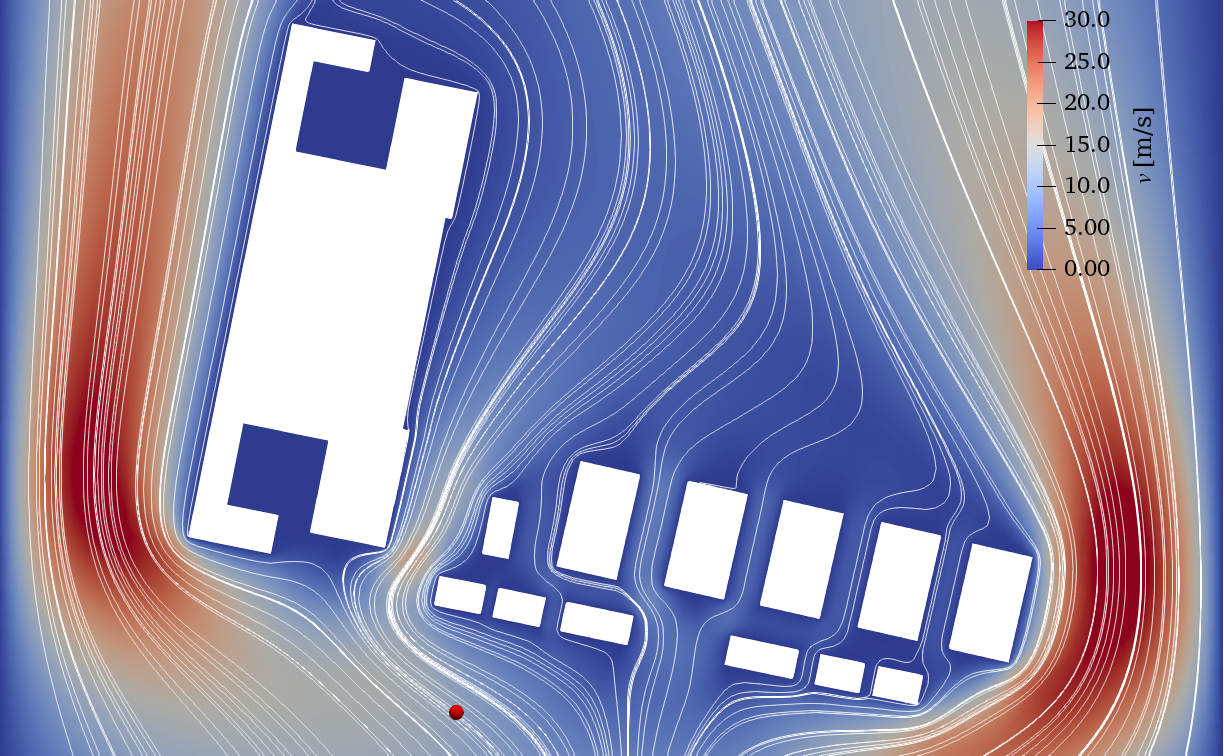}}
	\hspace{5pt}
	\subfloat[][Exemplary wind field of Geometry 2 for $\Reynolds\approx200$.
	\label{subfig:wind_field_henkel}]
	{\includegraphics[width=\columnwidth]{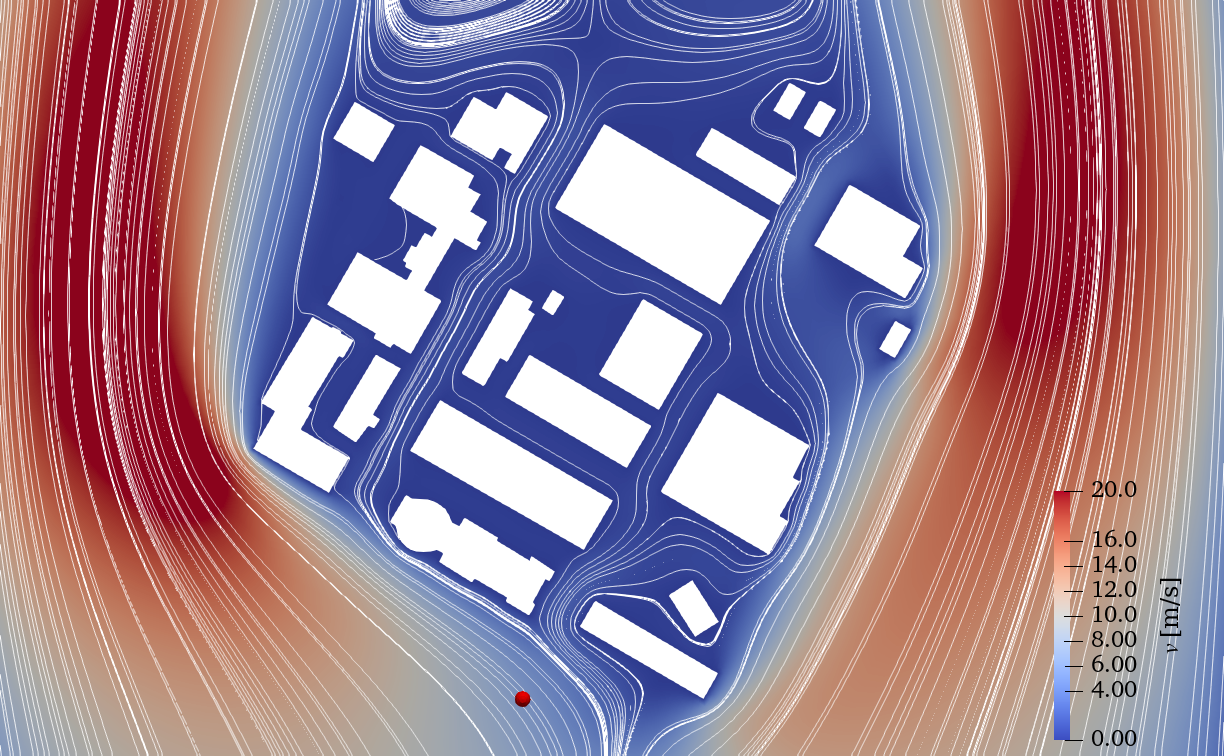}}\\
    \subfloat[][Concentration values right after the event.\label{subfig:conc_field_campus}]
	{\includegraphics[width=\columnwidth]{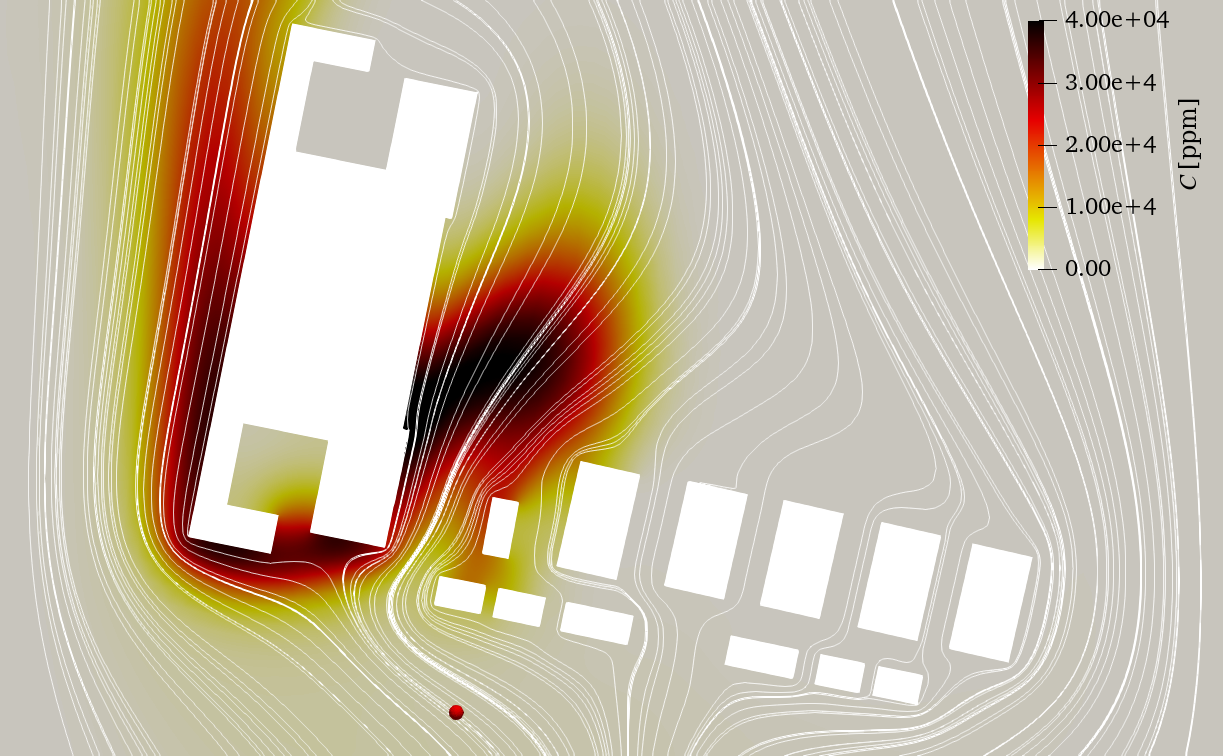}}
    \hspace{5pt}
    \subfloat[][Concentration values at the end of time analysis window.\label{subfig:conc_field_henkel}]
	{\includegraphics[width=\columnwidth]{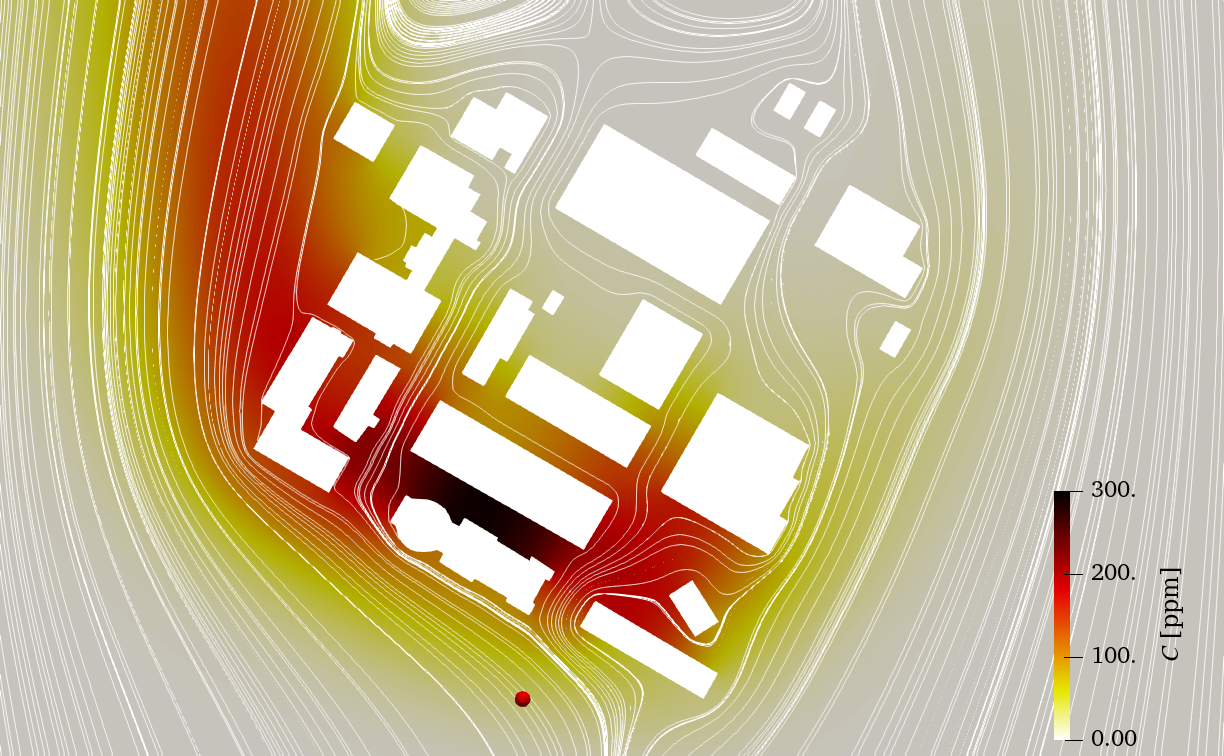}}
	\caption{Qualitative results for the two use-cases in terms of wind vector field and concentration distribution in [ppm].}
	\label{fig:qual_res}
\end{figure*}

To showcase the capabilities of our model, two different arrangements of real built environments with different levels of complexity are chosen as numerical examples. The two geometries are both obtained from OSM via an OverPy query, 
asking for all the buildings located within a rectangle bounded by minimum and maximum values of latitude and longitude. The first setup, referred hereinafter as \emph{Geometry 1}, reproduces a portion of the campus of the University of the Bundeswehr Munich, Germany, and contains a moderate number of buildings; the second set, \emph{Geometry 2}, entails a higher number of buildings and coincides with a portion of a chemical plant located in D\"{u}sseldorf, Germany.

\subsection{Analysis of computed wind and concentration fields}
The building arrangements for both setups can be observed in Fig.~\ref{fig:qual_res}. In both examples a Neumann boundary condition is applied on the outflow boundary. Moreover, all sides except for inflow and outflow sides are considered no-slip walls with imposed zero velocity. To account for different incoming wind conditions, a parametrized inhomogeneous Dirichlet boundary condition is imposed on the inflow side, resulting in different values of the Reynolds number $\Reynolds$ as shown in Tab.~\ref{tab:Parameters_MOR}.
Figures~\ref{subfig:wind_field_campus} and~\ref{subfig:wind_field_henkel} illustrate a qualitative representation of the numerical solution of the wind field for the two examples with $\Reynolds \approx 10$ and $\Reynolds \approx 200$, respectively. The vector field is represented by means of streamlines, while a background contour plot defines the wind magnitude. A qualitative analysis of the streamline pattern allows to verify the plausibility of the evolution of the gas contaminant distribution.

Unlike wind distribution, the AD phenomenon is inherently transient. To this end, an analysis time window is defined, lasting from $t_0=0\,\si{\second}$ to $t_\mathrm{f}=5\,\si{\second}$ for Geometry 1 and to $t_\mathrm{f}=50\,\si{\second}$ for Geometry 2, with constant time increments in both cases.
We study a scenario in which accidents (e.g. explosions) just before $t_0$ lead to an initial distribution of gas $c_0(\bx)$ in the form of a truncated Gaussian bell. The centers of the initial contaminant concentrations are located at the points highlighted by the coloured mark in Fig.~\ref{fig:qual_res}. Figure~\ref{subfig:conc_field_campus} shows the concentration distribution right after the event, where very high values, in the range of $10^4\,$ppm can be observed, before the contaminant is rapidly blown away. On the other hand, the snapshot of Fig.~\ref{subfig:conc_field_henkel} is taken at $t = t_\mathrm{f}$, and despite that the final concentration is orders of magnitude lower, it stagnates at potentially critical levels in zones of the domain characterized by low air circulation, qualifying them as the most dangerous parts.

The depicted results confirm that the contaminant transport follows the streamlines of the wind field and respects the influence of the buildings.

\begin{table}
\caption{Parameters of the presented numerical examples.}
\begin{center}
\begin{tabular}{|c|c|c|c|}
\hline
\multicolumn{2}{|c|}{\textbf{Parameter}} &\multicolumn{2}{|c|}{\textbf{Numerical Examples}} \\
\cline{3-4}
\multicolumn{2}{|c|}{ }& \textbf{\textit{Geometry 1}} & \textbf{\textit{Geometry 2}} \\
\hline
\multicolumn{2}{|c|}{Reynolds number} & $Re \in \left[5.5,515\right]$ & $Re \in \left[10,380\right]$\\ 
\hline
\multicolumn{2}{|c|}{Wind direction} & \multicolumn{2}{|c|}{South} \\ \hline
Subspace & POD ($\romdim$)& \multicolumn{2}{|c|}{10} \\ \cline{2-4}
dimension & DEIM ($\deimdim$) & \multicolumn{2}{|c|}{20} \\ \hline
\multicolumn{2}{|c|}{Snapshots (POD, DEIM)} & \multicolumn{2}{|c|}{50} \\ \hline
\multicolumn{2}{|c|}{Test set size (POD, DEIM)} & \multicolumn{2}{|c|}{20} \\ \hline
\end{tabular}
\label{tab:Parameters_MOR}
\end{center}
\end{table}

\subsection{Quantitative performance analysis of the reduced-order model}
To approach real-time applications, a good performance of the ROM
procedure is essential and therefore analyzed in the following. The
numerical parameters of the reduction and hyper-reduction are stated in
Tab.~\ref{tab:Parameters_MOR} along with the Reynolds number ranges
corresponding to the inflow velocity considered as parameter in the ROM.
Both the speed up and the error are evaluated over an independent test
set throughout the parameter space.

\begin{figure*}
	\centering
    \subfloat[][Normalized eigenvalues of snapshot correlation matrix of
    POD procedure.\label{subfig:eigs_combined}]
	{\includegraphics[width=\columnwidth]{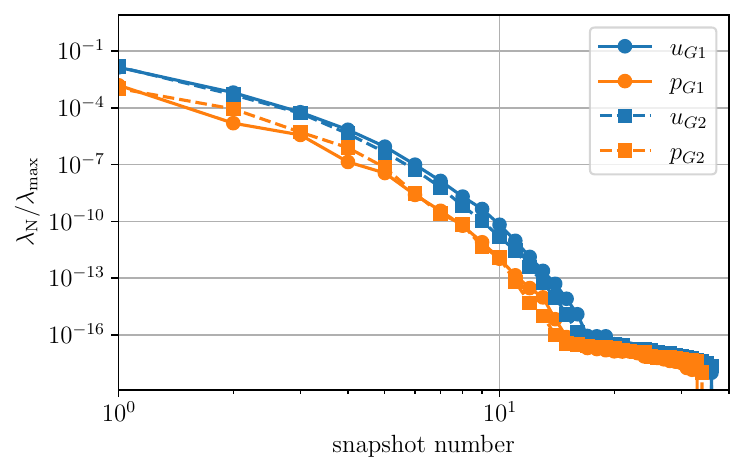}}
	\hspace{6pt}
	\subfloat[][Maximum relative error between ROM and FOM solutions.
	\label{subfig:mor_error_combined}]
	{\includegraphics[width=\columnwidth]{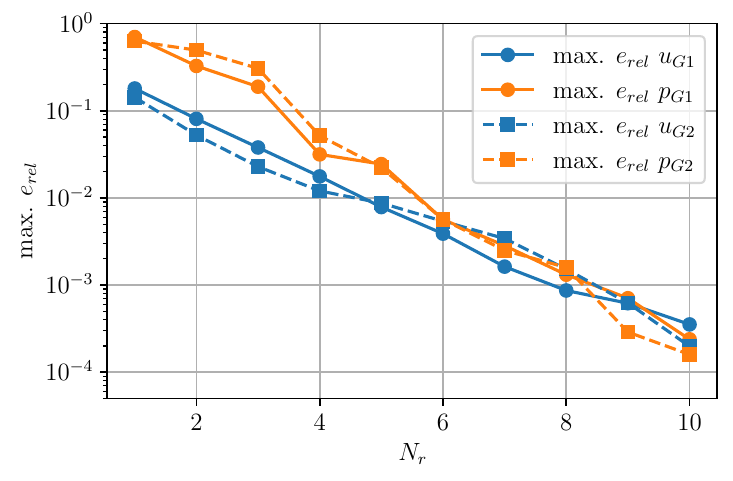}}\\
    \subfloat[][Speed up for Geometry 1 as function of the RB dimensions $N_r$.\label{subfig:speedup_campus}]
	{\includegraphics[width=\columnwidth]{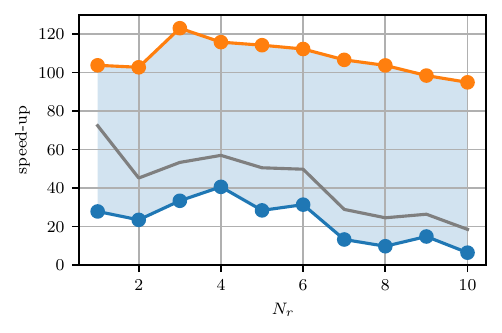}}
    \hspace{6pt}
    \subfloat[][Speed up for Geometry 2 as function of the RB dimensions $N_r$.\label{subfig:speedup_henkel}]
	{\includegraphics[width=\columnwidth]{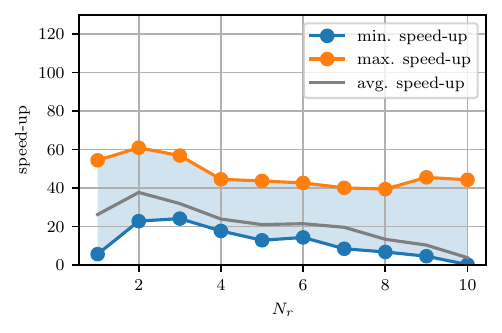}}
	\caption{Performance analysis of the reduced-order model (ROM) for both geometries.} 
	\label{fig:mor_campus_and_henkel}
\end{figure*}

Figure~\ref{fig:mor_campus_and_henkel} collects the ROM performance
results of both test cases. The quick decay of the eigenvalues in the
snapshot data correlation matrix shows that about 20 RB functions are
sufficient to cover the entire system characteristics for both
applications. Further evidence comes from the examination of the maximum
relative error over the entire computational domain and the entire
parameter space, which is shown in Fig.~\ref{subfig:mor_error_combined}
as a function of the RB dimension. The error is acceptable for small
dimensions of the RB ($\romdim < 10 $). When aiming for a maximum
deviation from the FOM results of less than $1\%$, a RB dimension of
$\romdim = 6$ is already enough.

Figure~\ref{subfig:speedup_henkel} shows the speed up of the ROM
evaluation in comparison to the FOM. As an overall trend, the speed up
decreases as the model complexity increases, as can be expected.
However, by choosing $\romdim = 6$ as explained above, a minimum,
average, and maximum speed up of approximately $30$, $50$, and $110$ is
achieved, on Geometry 1, for the entire parameter space. Also for the
more complex Geometry 2 a significant average speed up of $20$ is
obtained for $\romdim = 6$. In summary, the POD-based ROM procedure has
the potential to reduce the computational cost of urban physics
simulations and represents a first step towards real-time simulations.

\section{Conclusion and Outlook}\label{sec:conclusion}
As urban populations grow and critical infrastructures become
increasingly threatened, safeguarding against potential disruptions
becomes more significant than ever before. Among the possible threats,
dangerous gas contaminants dispersion stands out due to its potentially
catastrophic impact. The emergence of the DT framework offers promise in
analyzing and predicting such events, fostering situational awareness
and crisis management. This study proposes a computational framework for
analyzing airborne contaminant dispersion, leveraging on the automatic
generation of computational domains and solution processes. Employing
FEM-based CFD to obtain the numerical solutions of the systems of
equations governing the phenomena under examination, this approach
appears suitable in improving situational awareness and real-time
decision support within, e.g., evacuation scenarios.

The framework was tested with wind field and concentration computations
on two real world geometries. In both examples, an explosion-like
accident with an instant leakage was studied. However, with little
additional effort a continuous source can be considered as well within
the same framework. Furthermore, the developed workflow is easily
extended to further urban physics problem classes, such as simulating
the temperature field in built environments. Intrusive MOR methods were
explored to enhance computational efficiency and obtain a quantitative
assessment on the achievable accuracy levels. While the solution
accuracy was found to be sufficient, the speed up needs further
improvement.

The development and extension of the current virtual replica involves
further necessary steps to add to the model necessary features. Amongst
them, the most important is the substitution of hard-coded inputs with
real-time data coming from sensors and detection control units. The
increased simulation time associated to a continuous data stream will be
also carefully investigated and optimized, being this a potential
bottleneck.
Given the proposed scope of performing accurate real-time simulations,
MOR will be extended to the transient part of the solution step as well,
and the overall time required by a single time step will be analyzed to
guarantee the absence of over-run and lags. These topics will be
thoroughly investigated in future research and their features added to
the current model, aiming at the declared final goal of instantiating a
functional DT.

\section*{Acknowledgment}

The authors would like to acknowledge their colleagues at the department
of Digital Twins for Infrastructures at the DLR Institute for the
Protection of Terrestrial Infrastructures for  fruitful discussions and
helpful insights into DT coupling strategies.

\bibliographystyle{IEEEtran} 
\bibliography{bibliography}

\end{document}